\documentclass[aps,prx,twocolumn,superscriptaddress,showpacs]{revtex4-1}
\usepackage{amsmath}
\usepackage{amsfonts}
\usepackage{amssymb}
\usepackage{graphicx}
\usepackage{bm}
\usepackage{bbold}
\usepackage{float}
\usepackage{hyperref}
\usepackage{color}

\begin{document}
\title{Tricritical Ising phase transition in two-ladder Majorana fermion lattice}
\author{Xiaoyu Zhu}
\affiliation{National Laboratory of Solid State Microstructures and Department of Physics, Nanjing University, Nanjing 210093, China}
\affiliation{Department of Physics and Astronomy and 
Quantum Matter Institute, University of British Columbia, Vancouver, British Columbia, Canada V6T 1Z4}

\author{M. Franz}
\affiliation{Department of Physics and Astronomy and 
Quantum Matter Institute, University of British Columbia, Vancouver, British Columbia, Canada V6T 1Z4}

\date{\today}

\begin{abstract}
We introduce a two-ladder lattice model with interacting Majorana fermions that could be realized on the surfaces of a topological insulator film. We study this model by a combination of analytical and numerical techniques and find a  phase diagram that features both gapless and gapped phases as well as interesting phase transitions including a quantum critical point in the tricritical Ising (TCI) universality class. The latter occurs at an intermediate coupling strength at a meeting point of a first-order transition line and an Ising critical line and is known to be described by a superconformal field theory with central charge $c =\frac{7}{10}$.  We discuss the experimental feasibility of constructing the model and tuning parameters to the vicinity of the TCI point where signatures of the elusive supersymmetry can be observed.
\end{abstract}

\pacs{}

\maketitle

\section{Introduction}

Tricritical Ising (TCI) phase transition occurs where an Ising phase transition line meets a first order transition line. At this critical point, three different phases become indistinguishable. The critical exponents also experience a dramatic change when the system goes from Ising critical line to the TCI point as the latter belongs to a different universality class \cite{Henkel_Conformal}. In conformal field theory (CFT) language, TCI CFT is the second unitary minimal model, with central charge $c={7\over 10}$  whereas the Ising CFT has $c = {1\over 2}$ \cite{Francesco_Conformal}.
Recent interest in physical realizations of TCI CFT stems from the fact that it is the simplest known CFT that exhibits supersymmetry, a special type of symmetry that interchanges  bosonic and  fermionic fields. This enigmatic property has been conjectured to cure many problems in high energy physics but remains experimentally unobserved \cite{Martin1997}. 
Technically, the  TCI point  can be described by a superconformal field theory \cite{Friedan_Conformal,Friedan_Superconformal,Qiu_Supersymmetry} which is developed by extending the Virasoro algebra in CFT to its supersymmetric counterpart. 

There are two well-known spin models in the literature that realize TCI CFT: Blume-Capel model \cite{Blume_Theory,Capel_On} and Ising metamagnet model \cite{Henkel_The}. The former is a modified Ising model where each spin site is allowed to be vacant. The latter is an Ising antiferromagnetic model with next-nearest-neighbor ferromagnetic interaction. Despite extensive searches the TCI quantum criticality and the associated supersymmetry has yet to be observed experimentally in spin models.  Other condensed matter systems have been therefore  considered recently as potential platforms for the observation of TCI points and the associated supersymmetry.
Among these, condensed matter realizations of Majorana fermions -- Majorana zero modes (MZMs) \cite{Wilczek_Majorana,Alicea_New,Beenakker_Search,Stanescu_Majorana,Elliott_Colloquium} -- lend themselves naturally  to this task. A simple non-interacting 1D chain of MZMs (a critical Kitaev chain) maps onto the transverse field Ising model tuned to its critical point and realizes a $c={1\over 2}$ CFT \cite{Fendley2012}. Breaking the translation symmetry through dimerization produces gapped phases \cite{Kitaev2001} while coupling MZMs to bosonic modes \cite{Grover_Emergent} or adding four-fermion interactions \cite{Rahmani_Emergent,Rahmani_Phase} has been shown to give rise to TCI behavior with $c={7\over 10}$, by tuning a single model parameter.  
 
As the evidence supporting  the existence of MZMs in condensed matter systems has been growing \cite{Mourik_Signatures,Das_Zero-bias,Deng_Anomalous,Rokhinson_The,Finck_Anomalous,Nadj-Perge_Observation} the possibility of realizing the TCI criticality built on this platform is becoming more promising. With the rapid development in this field one can expect in the near future to manipulate MZMs and to be able to engineer interacting lattice models envisioned theoretically \cite{Grover_Emergent,Rahmani_Emergent,Rahmani_Phase}. At the same time, realization of these models presents significant challenges. The model of Ref.\ \cite{Grover_Emergent} requires Majorana modes to couple to bosonic (spin) modes and it is not clear how such coupling might be engineered and controlled. The model of Ref.\   \cite{Rahmani_Emergent} only requires short range 4-fermion interaction, which is generically present in the system, but TCI point occurs only at very strong coupling. 

In the present paper we construct a fermionic model with short range 4-fermion interactions that is more complex than models of Refs.\ \cite{Grover_Emergent,Rahmani_Emergent} but has an advantage that it exhibits a TCI point at weak to intermediate interaction strength and does not require coupling to bosonic modes. Similar to the spin models \cite{Blume_Theory,Capel_On,Henkel_The} two parameters must be tuned to reach the TCI point. The model can be realized in the Fu-Kane superconductor \cite{Fu2008} that occurs at the interface between a 3D strong topological insulator and an ordinary $s$-wave superconductor.  Here MZMs are bound in the cores of Abrikosov vortices and form a periodic lattice in the applied magnetic field. Effective theories for MZMs in such vortex lattices have been recently studied \cite{Chiu_Strongly,Pikulin_Interaction,Cobanera2015,Fu2015,Liu_Electronic,Murray2015} with the conclusion that they realize a convenient platform to probe interacting phases of Majorana fermions.

\section{The model and its realization}

\begin{figure}[t]
\includegraphics[scale = 0.29]{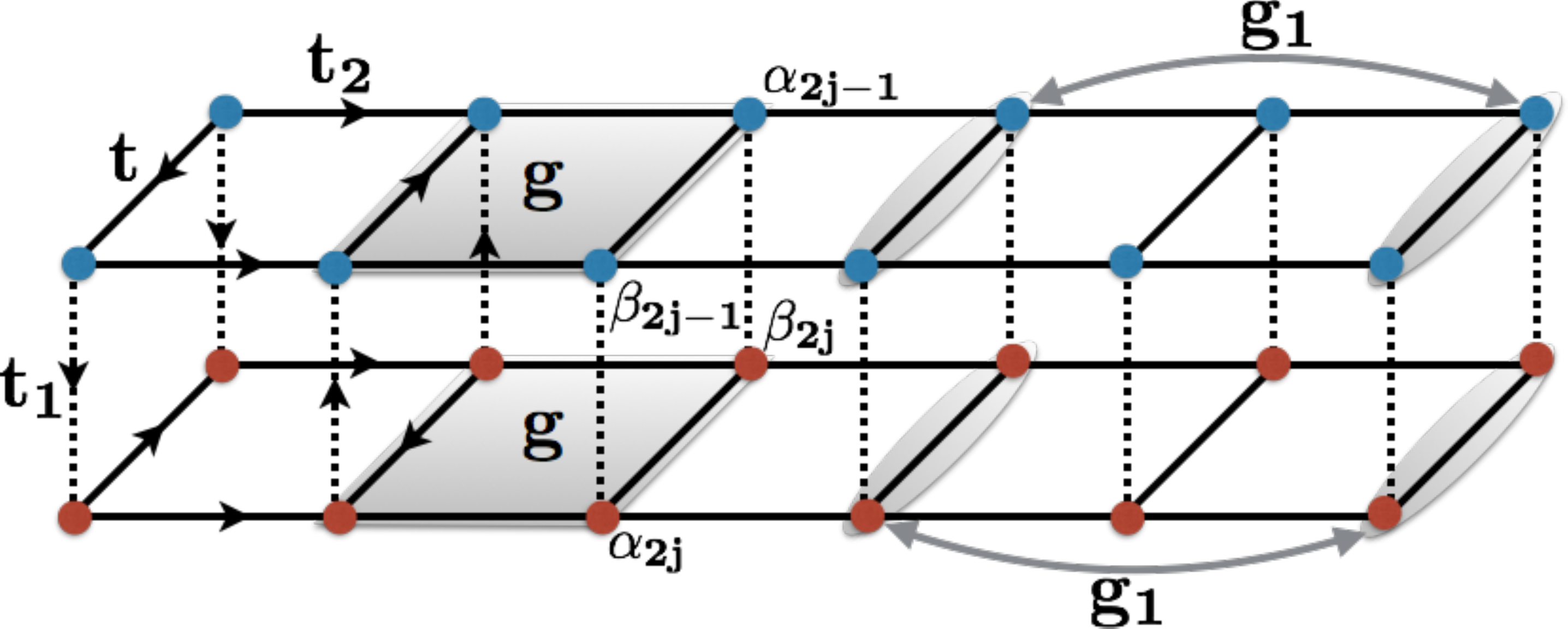}
\caption{Geometry of the two-ladder MZM lattice. Blue dots represent vortices and the red are antivortices. The arrow is added to indicate the sign of bilinear hopping terms. If the arrow starts from a site denoted by $\alpha$ to $\beta$, the bilinear term is  written as $i\alpha\beta$. The interaction term $g$ involves four MZMs on the nearest-neighbor rungs while $g_1$ connects  next-nearest-neighbor rungs.}\label{fig:lattice}
\end{figure}

The model we consider consists of two coupled ladders, each formed of  Majorana sites as schematically depicted in Fig.\ \ref{fig:lattice}. The upper ladder is composed of vortices, and the lower one of antivortices. Each of them hosts a single MZM. A realistic setup that can realize this model is a thin film of topological insulator in proximity of superconducting films on the two surfaces. When an external magnetic field perpendicular to the surfaces is applied vortices and antivortices are induced on the upper and the lower surface, respectively.  At the so called neutrality point (i.e. when the chemical potential $\mu$ of the TI coincides with the Dirac point), the Fu-Kane superconductor in each surface develops an extra chiral symmetry,\cite{Teo_Topological} which changes its non-interacting topological classification from $\mathbb{Z}_2$ to $\mathbb{Z}$. The latter implies that the bilinear tunneling terms between MZMs bounded in two vortices of the same type are prohibited while  4-fermion interaction terms are allowed and could thus be dominant. There is no restriction on the tunneling terms between a vortex and an antivortex. Slightly away from the neutrality point tunneling terms between two vortices of the same type are allowed but are generically small. In the two-ladder system we shall work in the regime where tunneling and interaction terms among vortices of the same type are comparable and the terms involving vortices and antivortices are dominated by tunneling.

If we assume that there are $2N$ MZM sites then the above discussion leads to the  following  Hamiltonian for the two-ladder system depicted in Fig.\ \ref{fig:lattice}
\begin{eqnarray}\label{eq:H}
H&=& \sum\limits_{j=1}^{N/2} (-1)^j [it(\alpha_{2j}\beta_{2j}+\alpha_{2j-1}\beta_{2j-1})\nonumber\\
      &+& it_1(\alpha_{2j-1}\beta_{2j} + \beta_{2j-1}\alpha_{2j})] + \sum\limits_{j=1}^{N-4}g_1\alpha_j\beta_j\alpha_{j+4}\beta_{j+4}\nonumber\\
      &+& \sum\limits_{j=1}^{N-2}[it_2(\alpha_j\alpha_{j+2} + \beta_j\beta_{j+2}) + g\alpha_j\beta_j\alpha_{j+2}\beta_{j+2}].
\end{eqnarray}
Here $\alpha_j$ and $\beta_j$ are MZM operators at site $j$, with $\{\alpha(\beta)_j,\alpha(\beta)_{j'}\}=2\delta_{j,j'}$.  The $t_1$ term couples MZMs bound in  vortices and antivortices, and the remaining terms all involve MZMs bound in the vortices of the same type. The sign of each bilinear term is chosen to satisfy the Grosfeld-Stern rule \cite{Grosfeld_Electronic}. We assume $g$ and $g_1$ to be positive, corresponding to attractive interactions. In principle, the signs of $t$, $t_1$ and $t_2$ can be either positive or negative but these are equivalent since one can always perform a unitary transformation that changes the sign of only one of them without affecting the others. For instance, to change the sign of $t$, we simply take $(\alpha,\beta)_j \rightarrow (\beta,\alpha)_j$, which has no influence on the energy spectrum. Hence we will only consider the case where all $t$'s and $g$'s are non-negative.  We further note that other tunneling and 4-fermion interaction terms are allowed by symmetries in the Hamiltonian (\ref{eq:H}) but these will not qualitatively change the conclusions reached below as long as their strength is not large. 

In general, one needs at least three independent parameters to realize a model capable of supporting a TCI point and in order to reach the transition two of them need to be fine tuned. Fermionic models with an extra symmetry \cite{Grover_Emergent,Rahmani_Emergent} represent an exception to this rule such that the TCI point is reached by tuning a single parameter. For our specific model, as we will see later, all the terms except $t_2$ in Eq.(\ref{eq:H}) are required to induce the TCI phase transition. To make it simple we begin by setting $t_2=0$ for the moment and recover it later to study how TCI points are affected. As mentioned earlier, a TCI point can be thought of as the meeting point of a first-order phase transition line and an Ising transition line. So the system is required to have both first-order transition points and Ising critical points. The first-order phase transition occurs naturally in our model when $t_1=0$. In this case the two ladders are decoupled and thus we need to only consider one of them. To see where the transition occurs we perform a unitary transformation, $(\alpha,\beta)_{2j-1} \rightarrow (\beta,\alpha)_{2j-1}$ for $j\in$ even, on the upper ladder and rewrite the Hamiltonian in terms of Dirac fermion operators $c_j = (\alpha_j + i\beta_j)/2$. We obtain
\begin{equation}\label{eq:Hf}
H_u = \sum\limits_{j\in\text{odd}}^{N}2n_j[t - 2g(1 - n_{j + 2}) + 2g_1(1 - n_{j + 4})],
\end{equation}
where $n_j = c^\dagger_j c_j$ is the occupation number operator that takes values either 0 (empty) or 1 (occupied). Note that constant terms are left out in Eq.\ (\ref{eq:Hf}) and periodic boundary condition, $n_{N + j} = n_j$, is assumed. 

Since $n_j$ is a good quantum number in Eq.\ (\ref{eq:Hf}), all the eigenstates of $H_u$ are also eigenstates of $n_j$, characterized by each site being empty or occupied. It is not difficult to see that the system favors a doubly degenerate ground state when $t < 2g$, with every other site being occupied. For $t>2g$  the ground state is unique with all sites empty. Evidently, the ground state experiences an abrupt change when the system passes the transition point at $t = 2g$. In addition when  $g_1>0$, the system at the transition point exhibits a finite gap $\Delta = 4g_1$ to the lowest excited state. Hence exactly at the transition point a level crossing occurs and the ground state is triply degenerate, which demonstrates a discontinuous phase transition. It should be noted that $g_1$ term is very important here: without it the ground state degeneracy at the transition point would be infinitely large in the thermodynamic limit, making it a multi-critical point instead of a first-order transition. This is why we need the $g_1$ term in our specific TCI model.

To identify the Ising phase transition line we restore $t_1$ and apply a different unitary transformation, $(\alpha,\beta)_{2j-1}\rightarrow (-\beta,\alpha)_{2j-1}$ and $(\alpha,\beta)_{2j}\rightarrow (\beta,\alpha)_{2j}$ for $j\in$ even, to the Hamiltonian in Eq.(\ref{eq:H}). We then translate it to spin language by performing a Jordan-Wigner (JW) transformation \cite{Fendley2012}
\begin{equation}
\alpha_j=\prod\limits_{k=1}^{j-1} \sigma^x_k\sigma^z_j,\beta_j=-\prod\limits_{k=1}^{j-1} \sigma^x_k\sigma^y_j,
\end{equation}
followed by a Kramers-Wannier duality transformation 
\begin{equation}
\sigma_j^x=\tau_{j-1}^x\tau_j^x,\ \ \sigma_j^z=\prod\limits_{k < j}\tau_k^z.
\end{equation}
The resulting Hamiltonian takes the form
\begin{eqnarray}\label{eq:HS}
H_S&=&\sum\limits_{j=1}^{N/2} -t(\tau_{2j-2}^x+\tau_{2j}^x)\tau_{2j-1}^x-t_1(1 + \tau_{2j-2}^x\tau_{2j}^x)\tau_{2j-1}^z\nonumber\\
&+&\sum\limits_{j=1}^{N/2-1} g(\tau_{2j-2}^x\tau_{2j}^x+\tau_{2j}^x\tau_{2j+2}^x)\tau_{2j-1}^x\tau_{2j+1}^x\nonumber\\
&-&\sum\limits_{j=1}^{N/2-2} g_1(\tau_{2j-2}^x\tau_{2j+2}^x+\tau_{2j}^x\tau_{2j+4}^x)\tau_{2j-1}^x\tau_{2j+3}^x,
\end{eqnarray}
where we defined $\tau_0^x = 1$. Clearly $\tau^x$ operators at even sites commute with the Hamiltonian above and thus represent conserved quantities. Inspecting each term in Eq.\ (\ref{eq:HS}) individually it is not difficult establish that the ground state favors the configuration in which the absolute values of the terms inside the parentheses are maximized. For instance, the ground state of the $g$ term would favor $\tau_{2j}^x\tau_{2j+2}^x = 1$ for all $j$ or else $\tau_{2j}^x\tau_{2j+2}^x = -1$ for all $j$. When considering the whole Hamiltonian we expect this argument to hold if all four terms favor the same configuration. Indeed such a configuration exists, having $\tau_{2j}^x = 1$ for all $j$. In this configuration it is easy to see that the remaining degrees of freedom are described by Ising metamagnet model well known to support the TCI point \cite{Henkel_The}. In the special case when $t$ and $g_1$ are absent the Hamiltonian becomes simply that of the transverse field Ising model in which the Ising phase transition occurs exactly at $t_1 = g$. With finite $g_1$, the transition point is expected to move towards a larger $t_1$ while preserving the Ising universality class.

\begin{figure}[hhh]
\includegraphics[scale = 0.94]{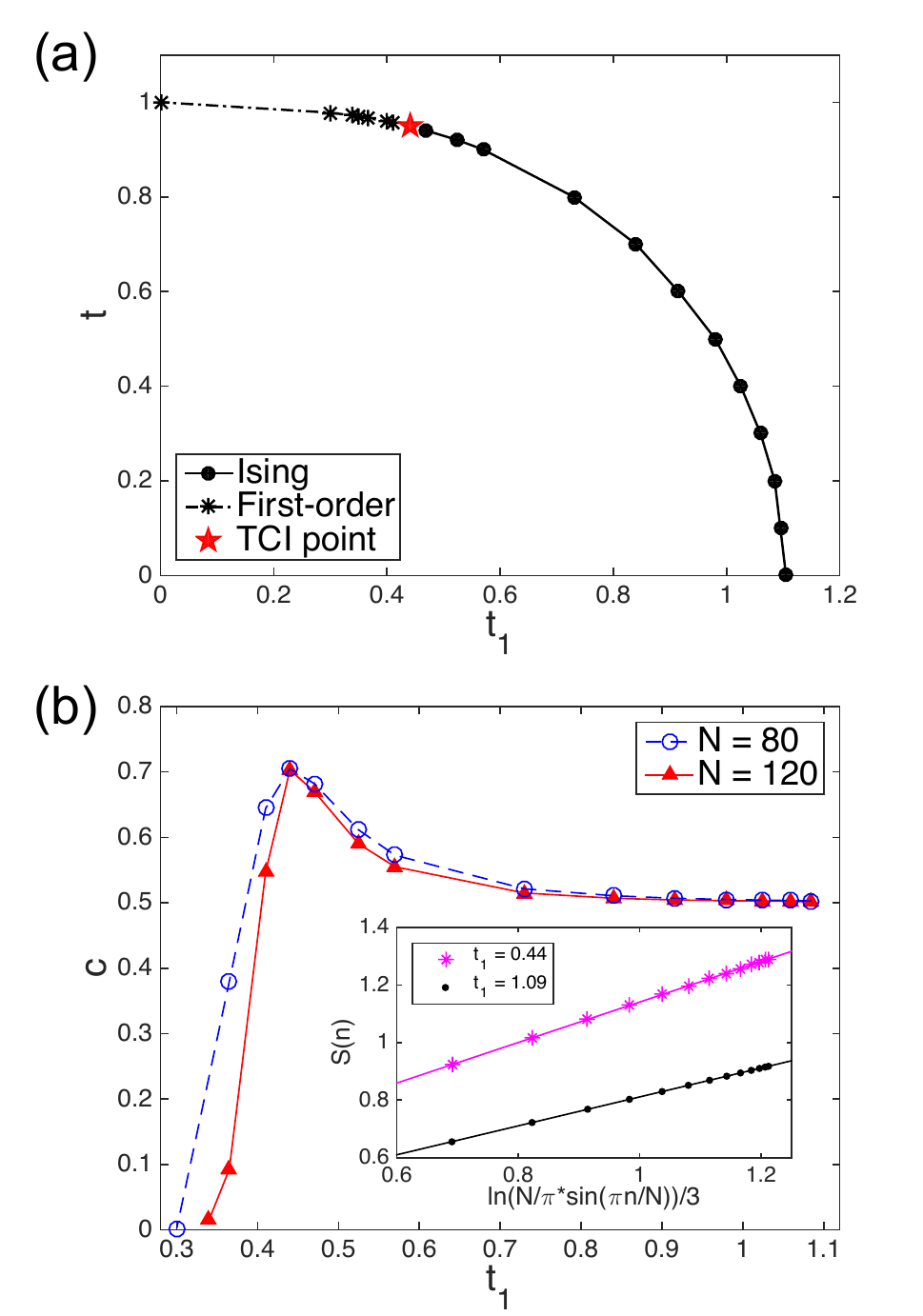}
\caption{a) Phase diagram of the model in the $t$-$t_1$ plane (the system size $N=120$) and b) central charge plot along phase transition line obtained by DMRG The inset shows the evolution of entanglement entropy with the length of subsystem, whose slope gives the central charge, according to Eq.(\ref{eq:central}). The  $t_1 = 0.44$ line corresponds to TCI point, with $c = {7\over 10}$, and the other line is in the Ising transition regime, with $c = {1\over 2}$. In both panels the interaction strength $g = 0.5$, and $g_1 = 0.4$ is assumed.}\label{fig2}
\end{figure}

\section{Numerical results}

The analytical arguments advanced in the previous Section  identified  first-order and Ising phase transitions in two special cases, each of which  corresponds to a point on one of the coordinate  axes in the $t$-$t_1$ phase diagram. In the Ising metamagnet model it is known that a phase transition line exists that connects these two points and a TCI point resides somewhere along this line.  On the basis of the above analysis we expect the same scenario to take place in our fermionic model. To verify the validity
of this conjecture and to locate the TCI point we performed extensive numerical analysis of the model defined by Eq.\ (\ref{eq:Hf}). Using the  density matrix renormalization group (DMRG) technique we located
the phase transition line as well as computed the central charge along the line. In the DMRG computations, we performed $10$-$12$ sweeps, and the truncation error in each sweep was set to be lower than $10^{-10}$. The central charge $c$ has been extracted from the fit to the entanglement entropy $S(n)$ of the ground state to the relation \cite{Calabrese_Entanglement},
\begin{equation}
S(n) = \frac{c}{3} \ln[\frac{N}{\pi}\sin(\frac{\pi n}{N})] + S_0,\label{eq:central}
\end{equation}
where $N$ is the system size, $n$ is the subsystem size and $S_0$ is a constant. It should be noted that Eq. (\ref{eq:central}) only works under (anti-)periodic boundary conditions. Representative results for generic values of $g$ and $g_1$ and $t_2=0$ are shown in Fig. \ref{fig2}. The central charge plot in Fig.\ \ref{fig2}(b) clearly demonstrates that a TCI point occurs characterized by a peak with $c = {7\over 10}$ which separates the discontinuous portion of the transition line ($c = 0$) from the Ising critical line ($c = {1\over 2}$). We note that the TCI points exist in the $t$-$t_1$ phase diagram whenever $g$ and $g_1$ are positive. Different values of these parameters simply alter  its position.

Until now we have not included the $t_2$ term which however generically will be present in any physical realization of the model. In the following we argue that weak to moderate values of $t_2$ (compared to $g$ and $g_1$) do not qualitatively change the phase diagram shown in Fig.\ \ref{fig2}(a) and, specifically the TCI point remains robustly present. We then present numerical evidence supporting these arguments. 

Consider first a case in which $t_2\gg t, t_1$. Noninteracting Hamiltonian (\ref{eq:Hf}) then describes four decoupled critical Majorana chains with only nearest-neighbor hopping. As is known, one such critical chain belongs to the Ising universality class with central charge $c = 1/2$. Four such chains together would trivially add and form a critical system with  $c = 2$. When interactions are switched on, we expect the gapless phase to persist for some range of coupling strengths followed by a phase transition to a gapped state. It is convenient to look at this in spin basis. Since in the absence of $t$ and $t_1$, the system decouples into two identical ladders we  need only to work with one of them. Again we perform a JW transformation on the upper ladder, and the resulting Hamiltonian reads
 \begin{equation}\label{eq:HP}
 H' =  \sum\limits_{j}-t_2(\sigma_j^z\sigma_{j+1}^z+\sigma_j^y\sigma_{j+1}^y)-g\sigma_j^x\sigma_{j+1}^x-g_1\sigma_j^x\sigma_{j+2}^x
 \end{equation}
which describes spin-$\frac{1}{2}$ Heisenberg XXZ model with next-nearest-neighbor interactions. To obtain Eq.\ (\ref{eq:HP}) we have applied a unitary rotation $\sigma_j^z\rightarrow -\sigma_j^y$ and $\sigma_j^y\rightarrow \sigma_j^z$ for $j\in$ odd, and the lattice index was relabelled since only the upper ladder is considered. It is obvious that a transition occurs when $t_2$ increases to $t_2 = g$ in the absence of $g_1$, at which point the system transits from a ferromagnetic phase to a gapless phase. Turning on finite $g_1$ is expected to move the transition point towards a larger $t_2$, since $g_1$ enhances the ferromagnetic order favored by the $g$ term. We can imagine that in the $t$-$t_1$ phase diagram the ferromagnetic phase would disappear when $t_2$ exceeds this transition point, and only gapped phases are left since non-zero $t$ and $t_1$ would gap out the $t_2$-dominated gapless phase. 
We thus conclude that in order to preserve the topology of the phase diagram indicated in  Fig.\ \ref{fig2}(a)  $t_2$ must not exceed the interaction parameters $g$ and $g_1$.  Otherwise, no TCI transition shall exist. 

To investigate the stability of the TCI point when the $t_2$ term is weak we actually only need to focus on the two special cases  of the model defined by $t_1=0$ and $t=0$. If the first-order and Ising transitions persist, we can expect the transition line connecting them to still exist and the TCI point thus to survive. 
\begin{figure}
\includegraphics[scale=0.94]{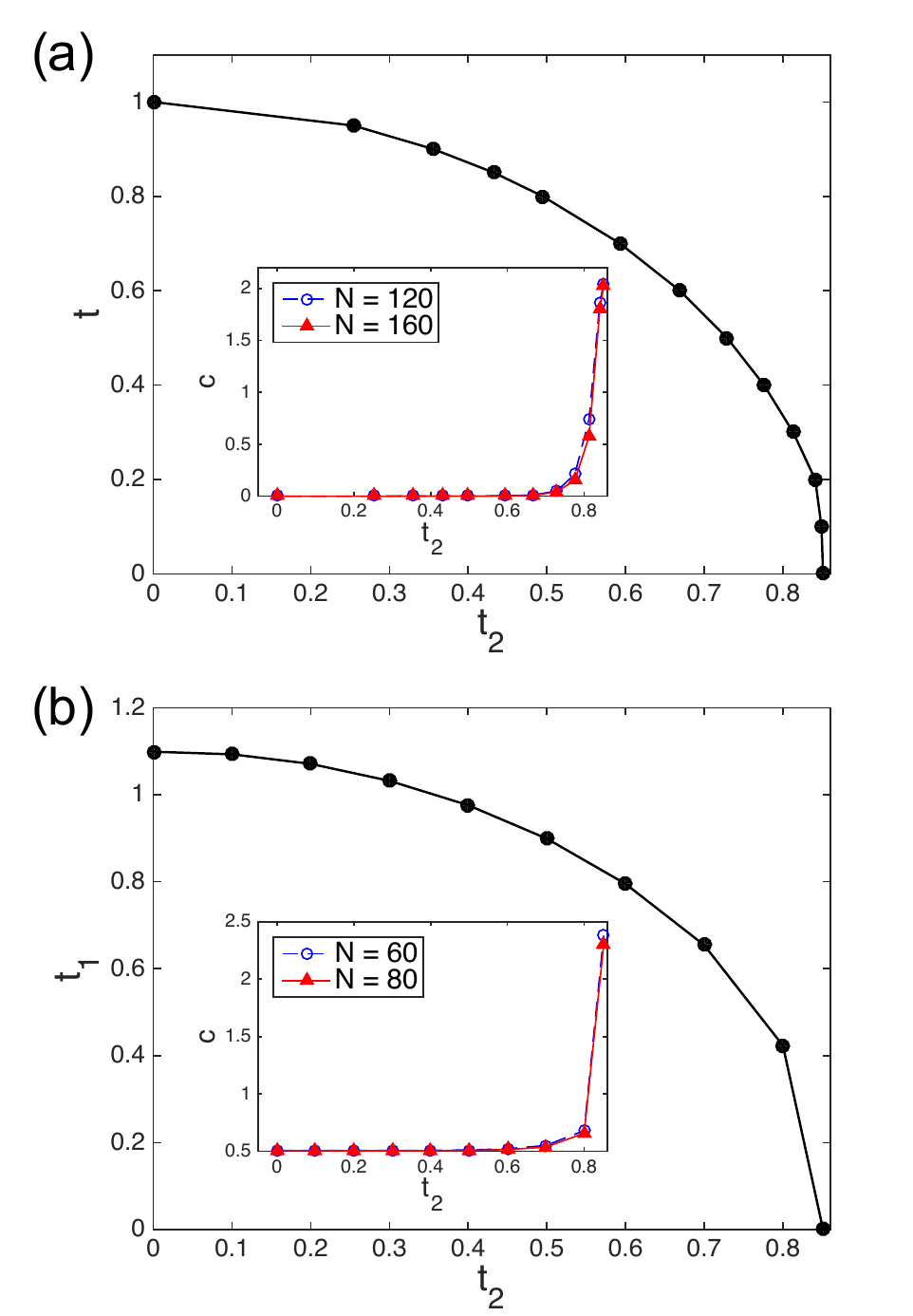}
\caption{Influence of $t_2$ on the first-order and Ising transitions in two special cases ($t_1 = 0$ for upper panel and $t = 0$ for the lower panel). As before  $g = 0.5$ and $g_1 = 0.4$. Small $t_2$ term does not affect the transition property as discussed in the text. The system size is $N=160$ for phase diagram in (a) and $N=80$ for that in (b)}\label{fig3}
\end{figure}
In the first case, a first-order transition occurs at $t = 2g$ in the absence of $t_1$. Since $t_2$ favors a gapless phase, one can expect increasing $t_2$ to drive the first-order transition towards a continuous one. However, for small $t_2$, the transition can still be discontinuous, considering the finite gap in the first-order transition point cannot be closed by an arbitrarily weak $t_2$. As indicated in Fig. \ref{fig3}(a), the first order transition point indeed continues moving towards small $t$ with the increase of $t_2$, until  in the vicinity of $t = 0$ the transition becomes continuous, characterized by $c = 2$. The same scenario unfolds in the second case, where an Ising transition occurs at $t_1=t_1^C(g,g_1)$ in the absence of $t$ and $t_2$. By turning on finite $t_2$, the critical point moves while still being of Ising type, as is shown in Fig. \ref{fig3}(b). We can thus expect the TCI point to still exist when $t_2$ is relatively small.  This is demonstrated in Fig. \ref{fig4} by explicit computation of the phase diagram for $t_2=0.2$. 

To reach the TCI transition point in the setup of Fig. \ref{fig:lattice}, therefore, we require the amplitude of the tunneling terms to be comparable with that of the four-fermion interaction terms, a regime of weak to intermediate coupling. The coupling between upper and lower ladder does not have to be strong compared to the intra-ladder hopping as can be seen from Fig. \ref{fig2}(a). Also we require that the bilinear coupling along the ladder ($t_2$ term) is comparable or weaker than the interaction strength.

\begin{figure}
\includegraphics[scale = 0.45]{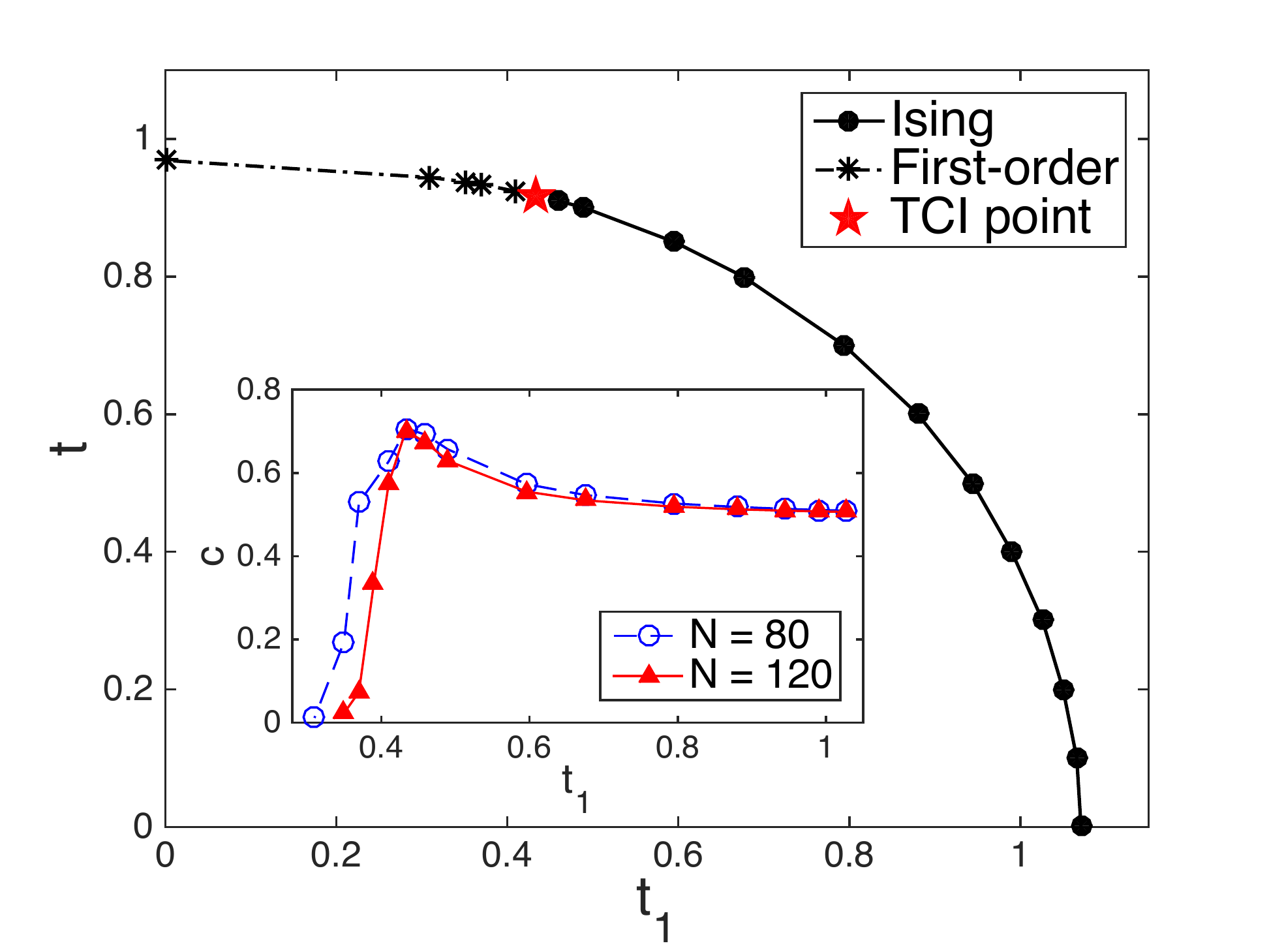}
\caption{Phase diagram ($N=120$) for nonzero $t_2=0.2$, and $g=0.5$, and $g_1= 0.4$. The TCI phase transition is stable against the inclusion of small $t_2$ except that the position of TCI point is slightly changed.}\label{fig4}
\end{figure}

%%%%%%%%%%%%%%%
\section{Summary and discussion}

To conclude, we constructed and analyzed a fermionic model with local interactions defined in a two-ladder Majorana lattice, which could be realized in a vortex lattice formed on two parallel surfaces of a proximitized topological insulator film. We established a connection (in a certain limit) between the low energy sector of our model and an Ising metamagnet spin model well known to support the tricritical Ising point. We have numerically verified the existence of the TCI point in our fermionic model and demonstrated that it occurs in the regime of weak to moderate interaction strength.

To reach the TCI point two model parameters must be independently tuned, similar to the spin models where TCI point is known to occur \cite{Blume_Theory,Capel_On,Henkel_The}. While such a tuning is unlikely to be practical in a spin system (since the coupling constants are typically fixed by the material parameters) it is conceivable, at least in principle, to do this in our model when realized as described above. Here, three quantities at least can be independently controlled: the topological insulator thickness $d$ (by fabrication), its chemical potential $\mu$ (by crystal chemistry and gating) and the applied magnetic field $B$.  These three quantities have very different effect on the system parameters: thickness $d$ for instance  affects predominantly the tunneling amplitude $t_1$ while the chemical potential influences mostly $t$ and $t_2$. Magnetic field in turn affects all parameters except $t_1$.  Therefore, by judicious choice of these inputs one could conceivably locate the transition line present e.g.\ in the $t$-$t_1$ plane and then move along the line to reach the TCI point. As suggested in Refs.\   \cite{Rahmani_Emergent,Rahmani_Phase} various phases and phase transitions can then be probed by tunneling (using scanning tunneling microscope for instance) into the zero mode states in the vortex cores where the tunneling conductance $G(V)=dI/dV$ exhibits a characteristic voltage dependence. For instance gapped phases of the model would show exponentially activated  behavior while Ising and TCI points exhibit power law $G\sim|V|^\alpha$ with $\alpha=0,2/5$, respectively. Supersymmetry will be most easily observable on the first order transition line in the close vicinity of the TCI point. Here it implies the existence of fermionic and bosonic excitations at the same energies. Fermionic excitations can be probed by an ordinary single-electron tunneling while bosonic excitations could be probed by pair tunneling with a superconducting tip \cite{Rahmani_Emergent,Rahmani_Phase}.  

We close by noting that ingredients necessary to start exploring various interacting models with Majorana zero modes, including the one introduced in this work, are currently in place.  Superconducting order has been induced in topological insulator  surfaces by multiple groups and in several different materials \cite{Kor11, Sac11, FQu12, Wil12, Cho13, YXu14, Zha14, PXu14,Harlingen2014}. The ability to tune the  chemical potential  to the vicinity of the Dirac point, required to bring in the regime with significant interaction strength, has also been demonstrated \cite{Cho13, YXu14, Zha14}. More recently, individual vortices have been imaged in these systems \cite{PXu14} and spectroscopic evidence indicative of MZMs in the cores of  vortices has been reported \cite{PXu15}. The current proposal requires inducing superconducting order on two surfaces of a topological insulator film which presents an additional experimental challenge. The rapid progress the field has been experiencing suggests that this challenge can be met in a not too distant future.

%%%%%%%%%%%%%%%
\section{Acknowledgments}
The authors are indebted to I. Affleck and A. Rahmani for illuminating discussions. DMRG simulations in this work were performed with ITensor library and we acknowledge M. Stoudenmire for helpful discussions on this. This work was supported by NSERC, CIfAR, Max Planck-UBC Centre for Quantum Materials, and China Scholarship Council (XZ).

\bibliography{tci.bib}

\end{document}